\documentclass[journal]{IEEEtran}
\usepackage{latexsym}
\usepackage{amsmath}
\usepackage{amssymb}
\usepackage{graphicx}

\def \F {{\textbf{F}}}
\newtheorem{lemma}{Lemma}
\newtheorem{remark}{Remark}
\newtheorem{theorem}{Theorem}
\newtheorem{corollary}{Corollary}
\newtheorem{definition}{Definition}
\newtheorem{proposition}{Proposition}
\newtheorem{example}{Example}

\title{Asymptotic Bound on Binary Self-Orthogonal Codes (confirmation no 18333)}

\author{Yang Ding \thanks{The work of Y. Ding was supported by the China Scholarship Council.
}
\thanks{Y. Ding is with the Department of Mathematics, Southeast
University, Nanjing, 210096, People's Republic of China (e-mail:
PG23067461@ntu.edu.sg). The work of Y. Ding was carried out while
the author was studying in Division of Mathematical Sciences,
School of Physical and Mathematical Sciences, Nanyang
Technological University, Singapore under the exchange program.}
\thanks{}}

\date{}
\begin{document}

\maketitle

%\title{\bf Access Structures of Elliptic Secret Sharing Schemes\footnote{This work of the first author
%was supported in part by  NNSF, China under Grant 90607005 and
%Distinguished Young Scholar Grant 10225106. The work of the second
%author was partially supported by Singapore MOE-AcRF Tier 2
%Research Grant T206B2204. The work of the third author was
%supported by Singapore MOE-ARF Research Grant R-146-000-029-112
%and National Scientific Research Project 973 of China
%2004CB318000.} }

\begin{abstract}We present two
constructions for binary self-orthogonal codes. It turns out that our constructions yield a constructive bound
on binary self-orthogonal codes. In particular, when the information rate $R=1/2$, by our constructive lower
bound, the relative minimum distance $\delta\approx 0.0595$ (for GV bound, $\delta\approx 0.110$). Moreover, we
have proved that the binary self-orthogonal codes asymptotically achieve the Gilbert-Varshamov bound.
\end{abstract}


\section{{\protect I}{\protect\small NTRODUCTION}}
In coding theory, we are interested in good codes with large
length, i.e., we want to find a family of codes with length
tending to $\infty $. For a
family of linear $[n,k,d]$ codes over $\F_{q}$, the ratio $R:=\underset{%
n\rightarrow \infty }{\lim }k/n$ and $\delta:
=\underset{n\rightarrow \infty }{\lim }d/n$ denote the information
rate and the relative minimum distance, respectively, of the
codes. The set $U_{q}\subseteq \lbrack 0,1]\times \lbrack 0,1]$
which is defined as follows: a point $(\delta ,R)\in
\mathsf{\mathbb{R}}^{2}$ with $0\leq \delta \leq 1$ and $0\leq
R\leq 1 $\ belongs to $U_{q}$ if and only if there exists a
sequence $\left\{
C_{i}=[n_{i},k_{i},d_{i}]\right\} _{i\geq 0}$ of codes over $\F_{q}$ such that%
\begin{equation*}
n_{i}\rightarrow \infty \text{, }\frac{d_{i}}{n_{i}}\rightarrow
\delta \text{ and }\frac{k_{i}}{n_{i}}\rightarrow R,\text{ as
}i\rightarrow \infty .
\end{equation*}%
A main coding problem is to determine the domain $U_{q}.$ Manin and Vl\u{a}du%
\c{t} gave a description of $U_{q}$ through a function $\alpha _{q}:[0,1]\rightarrow \lbrack 0,1]$ which is
defined by
\begin{equation*}
\alpha _{q}(\delta )=\sup \left\{ R:(\delta ,R)\in U_{q}\text{,
for }\delta \in \lbrack 0,1]\right\} \text{.}
\end{equation*}%
It is well-known that the function $\alpha _{q}$ is continuous and
decreasing, see \cite{tsfa}.

An $[n,k]$ linear code $C$ over the finite field $\F_{q}$ is a
linear $k$-dimensional subspace of $\F_{q}^{n}$. The dual code
$C^{\perp }$ of $C$ is defined as the orthogonal space of $C$,
i.e.,
\begin{equation*}
C^{\perp }=\{\mathbf{y}\in \F_{q}^{n}\text{ }|\text{ }\mathbf{xy}=0\text{ for every }\mathbf{x}\in C\},
\end{equation*}%
where $\mathbf{xy}=x_{1}y_{1}+x_{2}y_{2}+\cdots +x_{n}y_{n}$ is
the ordinary scalar product of vectors
$\mathbf{x}=(x_{1},x_{2},\cdots
,x_{n})$, $\mathbf{y}=(y_{1},y_{2},\cdots ,y_{n})$ in $\F_{q}^{n}.$ A code $C$ is self-othogonal if $C\subseteq C^{\perp }$, and self-dual if $%
C=C^{\perp }$. It is well-known that there exists a class of long binary self-dual codes which meet the
Gilbert-Varshmov bound \cite{mac2}. We employ the method which mentioned in \cite{mac2}, proof that binary
self-orthogonal codes also achieve the Gilbert-Varshmov bound. However, this result is not constructive.

To obtain the constructive bound on $R$ and $\delta$, we involved two different ways to construct binary
self-orthogonal codes. Both of the two constructions are based on a kind of algebraic geometry codes which
achieves the Tsfasman-Vl\v{a}du\c{t}-Zink bound. In the Construction A, we concatenate algebraic geometry codes
with binary self-orthogonal codes to obtain the desired codes. In the Construction B, we also get the desired
codes by considering self-orthogonal algebraic geometry codes and express these algebraic geometry codes into
binary self-orthogonal codes by employing the self-dual basis. Using these two constructions, we obtain a lower
bound on $R$ and $\delta$. In particular, using Construction B, we get $\delta\approx 0.0595$ when $R=1/2$ (by
Gilbert-Varshamov bound, $\delta\approx 0.110$ when $R=1/2$).

This correspondence is organized as follows. We first recall some basic results of concatenated codes,
Reed-Muller codes, Gilbert-Varshamov bound, and some well-known facts about algebraic geometry codes which are
necessary for our purpose. The main description of our two constructions are given in Section III, and we
calculate some examples. In Section IV, we have shown that there exists a binary self-orthogonal code achieving
the Gilbert-Varshamov bound. The conclusion of this paper is given in the last section.

\section{{\protect P}{\protect\small RELIMNARIES}}

In this section, we give some fundamental properties about concatenated codes, algebraic geometry codes and
Reed-Muller codes. We recall the results in \cite{mac}, \cite{stin} and \cite{xing} as follows.

Let $C$ be an $[s,v,w]$ code over $\F_{q^{k}}$ and, for $i=1,2,...,s$, let $%
\pi _{i}:\F_{q^{k}}\rightarrow \F_{q}^{n_{i}}$ be an
$\F_{q}$-linear injective map whose image $C_{i}=\mathrm{im}(\pi
_{i})$ is an $[n_{i},k,d_{i}]$ code over $\F_{q}$. The image $\pi
(C)$ of the following $\F_{q}$-linear injective
map:%
\begin{eqnarray}
\pi &:&C\rightarrow \F_{q}^{n_{1}+...+n_{s}}  \label{equ1} \\
\mathbf{c}=(c_{1},...,c_{s}) &\longmapsto &\pi(\mathbf{c})=(\pi _{1}(c_{1}),...,\pi _{s}(c_{s}))  \notag
\end{eqnarray}%
is an $[n_{1}+...+n_{s,}vk]$ linear concatenated code over
$\F_{q}.$

From the definition of the concatenated code, we know that if two codes $C$, $C'$ over $\F_{q^{k}}$ satisfy
$C\subseteq C'$, then the two concatenated codes $\pi (C)\subseteq \pi (C')$ over $\F_{q}.$
\begin{lemma}
If $\mathrm{im}(\pi _{i})=[n_{i},k,d_{i}]$ $(1\leq i\leq s)$ are
self-orthogonal codes, then $\pi (C)$ is also a self-orthogonal
code.

\begin{proof} Given any two codewords $\pi (\mathbf{c})$ $=(\pi _{1}(c_{1}),...,\pi _{s}(c_{s}))$ and $\pi
(\mathbf{c^{\prime }})=(\pi _{1}(c_{1}^{\prime }),...,\pi _{s}(c_{s}^{\prime }))$ of $\pi (C)$, where
$\mathbf{c}=(c_{1},c_{2},\cdots,c_{s})$ and $\mathbf{c}'=(c_{1}',c_{2}',\cdots,c_{s}')$ are two codewords of
$C$. Then
\begin{equation*}
\left(\pi (\mathbf{c}), \pi (\mathbf{c^{\prime }})\right)=\sum_{i=1}^{s}\left(\pi _{i}(c_{i}), \pi
_{i}(c_{i}^{\prime })\right),
\end{equation*}%
where $(,)$  stands for the ordinary scalar product over $\F_{q}$. Since $\mathrm{im}(\pi _{i})$ $(1\leq i\leq
s)$ are self-orthogonal, we have $\pi _{i}(c_{i})\cdot \pi _{i}(c_{i}^{\prime })=0$ for all $1\leq i\leq s.$
Thus $\pi (\mathbf{c})\cdot \pi (\mathbf{c^{\prime }})=0$, therefore, $\pi (C)$ is a self-orthogonal
code.\end{proof}
\end{lemma}
\begin{lemma} (\cite{mac})
Suppose the images $\mathrm{im}(\pi _{i})$ $(1\leq i\leq s)$ are identical and have parameters $[n,k,d].$ Then
$\pi (C)$ is an $[ns,vk]$ linear code over $\F_{q}$ with the minimum distance at least $wd$.
\end{lemma}

From now on, we assume that the images $\mathrm{im}(\pi_{i})$ $1\leq i\leq s$ are identical, and denote as
$\mathrm{im}(\pi_{\ast})$, i.e.,
$\pi((c_{1},c_{2},\cdots,c_{s}))=(\pi_{\ast}(c_{1}),\pi_{\ast}(c_{2}),\cdots,\pi_{\ast}(c_{s}))$ in equation
(1).

Next, we review some basic conclusions of algebraic geometry codes.

Let $X$ be a smooth, projective, absolutely irreducible curve of genus $g$ defined over $\F_{q}$, let
$\mathcal{D}$ be a set of $N$ $\F_{q}$-rational points of $X$ and let $G$ be an $\F_{q}$-rational divisor of $X$
such that $\mathrm{supp}(G)\cap \mathcal{D}=\emptyset$ and $2g-2<\mathrm{deg}(G)<N$, where $\mathrm{supp}(G)$
and $\mathrm{deg}(G)$ denote the support and the degree of $G$, respectively. Then the functional
algebraic-geometry code $C_{L}(G,\mathcal{D})$ with parameters $[N, \mathrm{deg}(G)-g+1, N-\mathrm{deg}(G)]$ can
be defined, see \cite{tsfa}.

Let $q=l^{2}$ be a square. It is known that there exists a family of algebraic curves $\{X_{i}\}$ over $\F_{q}$
with $g_{i}\rightarrow \infty $ attaining the Drinfeld-Vl\v{a}du\c{t} bound, i.e.,
\begin{equation*}
\lim_{i\rightarrow \infty }\sup (N(X_{i}/\F_{q})/g_{i})=l-1
\end{equation*}%
where $N(X_{i}/\F_{q})$ and $g_{i}$ are the number of $\F_{q}$-rational points and the genus of $X_{i}$,
respectively (see \cite{stin}). Then, the paper \cite{ch} constructs a family of algebraic geometry codes
$T_{i}=C_{L}(G_{i},\mathcal{D}_{i})=[N_{i},K_{i},D_{i}]_{q}$ achieving the Tsfasman-Vl\v{a}du\c{t}-Zink bound
where $\mathcal{D}_{i}$ contain all $\F_{q}$-rational point except only one rational point $P$ which is the
support of divisor $G_{i}$, i.e., we have
\begin{equation}
R_{1}+\delta _{1}=1-\frac{1}{l-1} \label{equ2}
\end{equation}
where%
\begin{equation*}
R_{1}\colon =\lim_{i\rightarrow \infty }\frac{K_{i}}{N_{i}}\text{
and } \delta _{1}\colon =\lim_{i\rightarrow \infty }\frac{D_{i}}{
N_{i}}.
\end{equation*}
denote the information rate and the relative minimum distance, respectively, of the codes.

For the Construction A, we also need some properties of Reed-Muller codes.

Let $\mathbf{v}=(v_{1},...,v_{m})$ denote a vector which ranges over $\F_{2}^{m}$, and $\mathbf{f}$ is the
vector of length $2^{m}$ by list of values which are taken by a Boolean function $f(v_{1},...,v_{m})$ on
$\F_{2}^{m}$.
\begin{definition} (\cite{mac})
The $r^{th}$ order binary Reed-Muller code (or RM code) $\mathfrak{R}(r,m)$ of
length $n=2^{m}$, for $0\leq r\leq m$, is the set of all vectors $\mathbf{f}$%
, where $f(v_{1},...,v_{m})$ is a Boolean function which is a
polynomial of degree at most $r.$
\end{definition}
\begin{lemma} (\cite{mac})
The $r^{th}$ order binary Reed-Muller code $\mathfrak{R}(r,m)$ has dimension $k=\sum_{i=0}^{r}\binom{m}{i}$ and
minimum distance $2^{m-r}$ for $0\leq r\leq m$, where $\binom{m}{i}$ are binomial coefficients.
\end{lemma}
\begin{lemma} (\cite{mac})
$\mathfrak{R}(m-r-1,m)$ is the dual code of $\mathfrak{R}(r,m)$ with respect to the ordinary scalar product, for
$0\leq r\leq m-1.$
\end{lemma}

From the definition of Reed-Muller codes, it is easy to known that we have $%
\mathfrak{R}(r_{1},m)\subseteq \mathfrak{R}(r_{2},m)$ when $0\leq
r_{1}\leq
r_{2}\leq m.$ By Lemma 4, when $r\leq \lfloor \frac{m-1}{2}\rfloor $, $%
\mathfrak{R}(r,m)$ is a self-orthogonal code. In particular, when
$m$ is an odd number, $%
\mathfrak{R}(\frac{m-1}{2},m)$ is a self-dual code.

Now, let $m$ go through all positive odd number, we get a family
of self-dual Reed-Muller codes $\mathfrak{R}(r,m),$ where $r=\frac{m-1}{2}%
,$ with parameters $[2^{m},2^{m-1},2^{\frac{m+1}{2}}]$.

At the end of this section, in order to compare our bound with the existed bound, we give the asymptotically
Gilbert-Varshamov bound.
\begin{lemma}
(Asymptotic Gilbert-Varshamov Bound) If $0\leq \delta \leq \frac{q-1}{q}$
then\begin{equation}\alpha_{q}(\delta)\geq 1-H_{q}(\delta ),  \label{equ3}\end{equation} where $H_{q}(\delta )$
is $q$-ary entropy function defined by

${\small H_{q}(x)=}{\small \left\{%
\begin{array}{ll}
    x\log_{q}(q-1)-x\log _{q}x-(1-x)\log _{q}(1-x), & \hbox{} \\
    \;\;\;\;\;\;\;\;\;\;\;\;\;\;\;\;0<x\leq (q-1)/q;& \hbox{} \\
    0,\;\; x=0. & \hbox{} \\
\end{array}%
\right.     }$\end{lemma}
\begin{remark} In Fig.1 we show this bound for $q=2$.\end{remark}

\section{{\protect C}{\protect\small ONSTRUCTIONS} {\small OF} {\protect S}{\protect \small ELF-ORTHOGONAL} {\protect C}{\protect\small
ODES}} In this Section, we will present two constructions of binary self-orthogonal codes.
\subsection{Construction A}
Assume that $q=2^{2t}$ in this subsection. Let $\mathrm{im}(\pi _{\ast})$ be an binary $[n,2t]$ linear code.

Let $T_{i}=[N_{i},K_{i},D_{i}]$ be a family of algebraic geometry codes over $\F_{2^{2t}}$ achieving the
Tsfasman-Vl\v{a}du\c{t}-Zink bound, i.e.,
\begin{equation}
R_{1}+\delta _{1}=1-\frac{1}{2^{t}-1} \label{equ4}
\end{equation}
where%
\begin{equation*}
R_{1}\colon =\lim_{i\rightarrow \infty }\frac{K_{i}}{N_{i}}\text{ and }%
\delta _{1}\colon =\lim_{i\rightarrow \infty }\frac{D_{i}}{N_{i}}.
\end{equation*}

Now we state our first construction.
\begin{proposition}
Let $C_{0}$ be a self-orthogonal code over $\F_{2}$ with parameters $[n,2t,d]$, take $C_{0}$ as
$\mathrm{im}(\pi_{\ast})$, concatenate the family of algebraic geometry codes $T_{i}$ and $C_{0}$ under the map
$\pi=(\pi_{\ast}, \pi_{\ast}, \cdots, \pi_{\ast})$, then we obtain a family of binary self-orthogonal codes
$C_{i}$ with parameters $[nN_{i},2tK_{i},dD_{i}]$. Moreover, we have asymptotic equation
\begin{equation}
R+\frac{2t}{d}\delta=\frac{2t}{n}(1-\frac{1}{2^{t}-1})
\label{equ5}
\end{equation}
where $R$ and $\delta$ denote the information rate and the
relative minimum distance, respectively, of the concatenated codes
$C_{i}$.

\begin{proof} The result follow immediately consequence of the properties of algebraic geometry codes $T_{i}$ and
concatenated codes.\end{proof}
\end{proposition}

Now we give some examples to illustrate the result in Proposition 1.

\begin{example} (RM codes) If we fixed an odd number $m\geq 3$, then we get a  binary
self-dual code [$2^{m},2^{m-1},2^{(m+1)/2}]$. Let $2t=2^{m-1}$, then $\F_{2^{2t}}=\F_{2^{2^{m-1}}}$.

It is well-known that there exists a family of algebraic geometry codes $T_{i}$ over $\F_{2^{2^{m-1}}}$ with
parameters $[N_{i},K_{i},D_{i}]$ satisfy the equation (4). Then by Proposition 1, we get a family of binary
concatenated codes $C_{i}=[2^{m}N_{i},2^{m-1}K_{i},2^{\frac{m+1}{2}}D_{i}]$. Asymptotically, we have the
equation
\begin{equation}R+2^{\frac{m-1}{2}}\delta=\frac{1}{2}(1-\frac{1}{2^{2^{m-2}}-1})\label{equ6}\end{equation}
Thus when we go through all odd number $m\geq 3$, we get a sequence of equations for $R$ and $\delta$.
\end{example}

\begin{example}(Some special binary self-orthogonal codes)
From $\cite{mac3}$ and $\cite{dod}$, we get several optimal self-orthogonal codes. Using these codes to do the
concatenation, we get some equations about $R$ and $\delta$ for small $t$. We list them in Table I. The last
column of Table I was calculated by (\ref{equ5}).

\begin{table}
  \centering
  \caption{Example 2}\label{tab1}
\begin{tabular}{|c|c|c|}
  \hline
  binary codes  & $t$ & equations for $R$ and $\delta$ \\
  \hline
  $[22,10,8]$ & $5$ & $R+\frac{5}{4}\delta=\frac{150}{341}$ \\

  $[24,12,8]$ & $6$ & $R+\frac{3}{2}\delta=\frac{31}{63}$ \\

  $[28,14,6]$ & $7$ & $R+\frac{7}{4}\delta=\frac{63}{127}$\\

  $[40,20,8]$ & $10$ & $R+\frac{5}{2}\delta=\frac{511}{1023}$\\

  $[44,22,8]$ & $11$ & $R+\frac{11}{4}\delta=\frac{1023}{2047}$ \\

  $[64,32,12]$ & $16$ & $R+\frac{8}{3}\delta=\frac{32767}{65535}$\\
  \hline
\end{tabular}
\end{table}
\end{example}

Using these two examples, we get an asymptotic bound for $\alpha_{2}(\delta)$.

\subsection{Construction B}

In this subsection, we will give another construction of binary self-orthogonal codes. Let us first recall the
definition of self-dual basis.

Let $\{e_{1},\cdots,e_{k}\}$ be an $\F_{q}$-basis of $\F_{q^{k}}$. A set $\{e_{1}',\cdots,e_{k}'\}$ of
$\F_{q^{k}}$ is called the dual basis of $\{e_{1},\cdots,e_{k}\}$ if we have\begin{equation*}
\mathrm{Tr}_{\F_{q^{k}}/\F_{q}}(e_{i}e_{j}')=\delta_{ij}=\left\{%
\begin{array}{ll}
    $0$, & \hbox{$i\neq j$;} \\
    $1$, & \hbox{$i=j$,} \\
\end{array}%
\right.    \end{equation*}(Kronecker symbol). It is well-known that the dual basis always exists. We say that a
basis is self-dual if it is its own dual. It is well-known that the self-dual basis always exists when
$\mathrm{char}(\F_{q})=2$.

Now we consider the finite field $\F_{2^{2t}}$, we know that there exists a self-dual $\F_{2}$-basis
$\{e_{1},\cdots,e_{2t}\}$ of $\F_{2^{2t}}$. Then for any element $\alpha$ in $\F_{2^{2t}}$, there exists a
unique $2t$-tuple vector $\mathbf{\alpha}^{(e)}=(\alpha_{1},\cdots,\alpha_{2t})\in \F_{2}^{2t}$ such that
$\alpha=\sum_{i=1}^{2t}\alpha_{i}e_{i}$. For any two elements $\alpha$ and $\beta$ of $\F_{2^{2t}}$, we have
\begin{equation*}\mathrm{Tr}_{\F_{2^{2t}}/\F_{2}}(\alpha\beta)=(\mathbf{\alpha}^{(e)},\mathbf{\beta}^{(e)})=\sum_{i=1}^{2t}\alpha_{i}\beta_{i},\end{equation*}
where $(,)$ stands for the ordinary scalar product over $\F_{2}$. Thus, we have a one-to-one correspondence
$\rho$ between $\F_{2^{2t}}^{n}$ and $\F_{2}^{2tn}$ such that
$\rho(\mathbf{a})=\rho((a_{1},\cdots,a_{n}))=(\mathbf{a_{1}}^{(e)},\cdots,\mathbf{a_{n}}^{(e)})$, where
$\mathbf{a_{i}}^{(e)}\; (1\leq i\leq n)$ is a vector of length $2t$ over $\F_{2}$ and
$\mathrm{Tr}_{\F_{2^{2t}}/\F_{2}}(\mathbf{a}\cdot
\mathbf{b})=\mathrm{Tr}_{\F_{2^{2t}}/\F_{2}}(\sum_{i=1}^{n}a_{i}b_{i})=\sum_{i=1}^{n}(\mathbf{a_{i}}^{(e)},\mathbf{b_{i}}^{(e)})$,
where $(\cdot)$ stands for the ordinary scalar product over $\F_{2^{2t}}$. Thus we have

\begin{lemma}Let $C$ be a self-orthogonal code over $\F_{2^{2t}}$,
then $\rho(C)$ is a self-orthogonal code over $\F_{2}$.

\begin{proof} For any two codewords $\rho(\mathbf{a})$ and $\rho(\mathbf{b})$ of $\rho(C)$
\begin{equation*}(\rho(\mathbf{a}),\rho(\mathbf{b}))=\sum_{i=1}^{n}(a_{i}^{(e)},b_{i}^{(e)})=\mathrm{Tr}_{\F_{2^{2t}}/\F_{2}}(\mathbf{a}\cdot\mathbf{b})=0
\end{equation*}
the last equality holds because $C$ is a self-orthogonal code. \end{proof}
\end{lemma}

To show our construction, we also need the result of self-orthogonal codes from \cite{stin2}:

\begin{lemma} (\cite{stin2}) Let $q=l^{2}$ be a square. Then the class of
self-orthogonal codes meet the Tsfasman-Vl\v{a}du\c{t}-Zink bound. More precisely, we have the following holds.
\begin{itemize} \item Let $0\leq R\leq 1/2$   and  $\delta\geq  0$  with $R=1-\delta-1/(l-1)$. Then there is a
sequence $(C_{j})_{j\geq 0}$ of  linear  codes $C_{j}$ over $\F_{q}$  with  parameters $[n_{j},k_{j},d_{j}]$
such    that the following:\end{itemize}
\begin{enumerate}
    \item all $C_{j}$ are self-orthogonal codes;
    \item $n_{j}\rightarrow \infty$ as $j\rightarrow \infty$;
    \item $\lim_{j\rightarrow \infty}k_{j}/n_{j}\geq R$ and $\lim_{j\rightarrow \infty}d_{j}/n_{j}\geq
    \delta$.
\end{enumerate}
\end{lemma}
\begin{remark} The existence of the self-orthogonal codes in Lemma 7 is constructive. For the detail of the construction of this codes, we refer to \cite{stin2}.\end{remark}

Then by Lemma 7, we know that there exists a class of
self-orthogonal codes over $\F_{2^{2t}}$ which meet the
Tsfasman-Vl\v{a}du\c{t}-Zink bound. Now, we give the
characterization of our construction.

\begin{proposition}
Let $C_{i}$ be a family of self-orthogonal codes over $\F_{2^{2t}}$ which meets the Tsfasman-Vl\v{a}du\c{t}-Zink
bound with parameters $[n_{i},k_{i},d_{i}]$, i.e.,
\begin{equation*}\lim_{i\rightarrow \infty}\left(\frac{k_{i}}{n_{i}}+\frac{d_{i}}{n_{i}}\right)=1-\frac{1}{2^{t}-1}.\end{equation*} Then $\rho(C_{i})$ is a family of
self-orthogonal codes over $\F_{2}$ with parameters
$[2tn_{i},2tk_{i},d_{i}]$. Moreover, we have equation
\begin{equation}R+2t\delta=1-\frac{1}{2^{t}-1}, \label{equ7}\end{equation}where
$R$ and $\delta$ denote the information rate and relative minimum distance, respectively, of the codes
$\rho(C_{i})$.
\end{proposition}

\begin{example} Using this construction, we get the equations of $R$
and $\delta$ in Table II. The second column of Table II was calculated by (\ref{equ7}). In particular, it is
easy to see that when we choose $t=3$ and $R=1/2$, we get the best value of $\delta$, $\delta\approx 0.0595$
from (7) (for asymptotic Gilbert-Varshamov bound, we have $\delta\approx 0.110$).

\begin{table}
  \centering
  \caption{Example 3}\label{tab2}
  \begin{tabular}{|c|c|c|}
    \hline
     $t$ & equations for $R$ and $\delta$& $R=1/2$ \\
    \hline
      2 & $R+4\delta=\frac{2}{3}$ & $\delta\approx 0.0417$ \\
      3 & $R+6\delta=\frac{6}{7}$& $\delta\approx 0.0595$ \\
      4 & $R+8\delta=\frac{14}{15}$ & $\delta\approx 0.05417$ \\
      5 & $R+10\delta=\frac{30}{31}$& $\delta\approx 0.04677$ \\
    \hline
  \end{tabular}
\end{table}
\end{example}

\section{\protect G{\protect\small ILBERT}-V{\protect\small ARSHAMOV}{\protect\ B{\protect\small OUND
}}} In this section, by mimicking the idea in [2], we give the proof that there exists a family of binary
self-orthogonal codes achieving the Gilbert-Varshamov bound. For binary self-orthogonal code, it is easy to know
that the weight of every codeword is even. Now we assume that the length of code $n$ is also an even number.

We first introduce two notations. Let ${\cal A} $ be the set of self-orthogonal codes of length $n$ over
$\F_{2}$, and let ${\cal A}_{1}$ denote the subset of ${\cal A}$ consisting of all self-dual codes of length $n$
over $\F_{2}$.

\begin{lemma}
(\cite{mac2}) Let $n=2h$ and, let $C$ be an $[n,s]$ binary self-orthogonal code. The number of codes in ${\cal
A}_{1}$ which contain $C$ is
\begin{equation*}
(2^{h-s}+1)(2^{h-s-1}+1)\cdots (2^{2}+1)(2+1).
\end{equation*}
\end{lemma}

Let $\sigma _{n,k,s}$, $s\leq k<h$, be the number of self-orthogonal codes $D$ with parameters $[n,k]$ which
contain the given code $C$. In the proof of Lemma 8, the authors establish a recursion formula for $\sigma
_{n,k,s}$.
\begin{equation}
\sigma _{n,k+1,s}=\sigma _{n,k,s}\times\frac{2^{n-2k}-1}{2^{k-s+1}-1}. \label{equ8}
\end{equation}
Then we have
\begin{corollary}
The number of codes in ${\cal A}$ of dimension $k$ is
\begin{equation}
\frac{(2^{n-2(k-1)}-1)(2^{n-2(k-2)}-1)\cdots (2^{n}-1)}{(2^{k}-1)(2^{k-1}-1)%
\cdots (2-1)}.  \label{equ9}
\end{equation}

\begin{proof} It is easy to know that every self-orthogonal code with dimension $k$ must contain the trivial
code ${\textbf{0}}$. Let $s=0$, then $\sigma_{n,k,0}$ is the number we require. Using the recursion formula we
get (\ref{equ9}).
\end{proof}
\end{corollary}

\begin{corollary}
Let $\mathbf{v}$ be a vector other than $\mathbf{0,1}$ with $\mathrm{wt}(\mathbf{v})\equiv0(\textrm{mod }2)$.
The number of codes in ${\cal A}$ of dimension $k$ which contain $\mathbf{v}$ is
\begin{equation}
\frac{(2^{n-2(k-1)}-1)(2^{n-2(k-2)}-1)\cdots (2^{n-2}-1)}{%
(2^{k-1}-1)(2^{k-2}-1)\cdots (2-1)}.  \label{equ10}
\end{equation}

\begin{proof} It is easy to know that every self-orthogonal code containing the vector $\mathbf{v}$ must contain the code
$C$, where $C$ is the linear code with basis $\{\mathbf{v}\}$. Then $\sigma_{n,k,1}$ is the number we require.
Using the recursion formula we get (\ref{equ10}).
\end{proof}
\end{corollary}

Using these two Corollaries, we have
\begin{theorem}
Let $r$ be a positive integer such that
\begin{equation}
\binom{n}{2}+\binom{n}{4}+\binom{n}{6}+\cdots +\binom{n}{2(r-1)}<\frac{%
2^{n}-1}{2^{k}-1}.  \label{equ11}
\end{equation}
Then there exists an $[n,k]$ self-orthogonal code with minimum
distance at least $2r.$

\begin{proof}The theorem is an immediate consequence of Corollaries 1 and 2.\end{proof}\end{theorem}
\begin{remark} For any $0\leq \delta\leq 1/2$, let $r=\lfloor\frac{\delta n}{2}\rfloor$, then\begin{equation*}
k=\left\lfloor\log_{2}\left(\frac{2^{n}-1}{\binom{n}{2}+\binom{n}{4}+\cdots+\binom{n}{2(r-1)}}\right)\right\rfloor\end{equation*}
satisfy (11), i.e., there exists an $[n,k,2r]$ binary self-orthogonal code and asymptotically, we
have\begin{equation} \frac{k}{n}\rightarrow 1-H_{2}(\delta).\end{equation} By Lemma 5, (12) implies that the
binary self-orthogonal code meets the Gilbert-Varshamov bound.\end{remark}

\section{{\protect C}{\protect\small %
ONCLUSION}}

Using these two constructions, we get a sequence of equations on $R$ and $\delta$. Then we get a constructive
bound on $\alpha_{2}({\delta})$ by combining the equations (5) and (7). We draw the figure of this bound in
Fig.1. When $R\rightarrow 0$, the constructive bound (5) is better than the constructive bound (7). When
$R\rightarrow 1/2$, the constructive bound (7) is better than the constructive bound(5). In Section IV, we proof
that binary self-orthogonal codes meet the Gilbert-Varshamov bound, we also show the figure of this bound for
self-orthogonal codes in Fig.1.

\begin{figure}
  % Requires \usepackage{graphics}
  \includegraphics[width=3.5in]{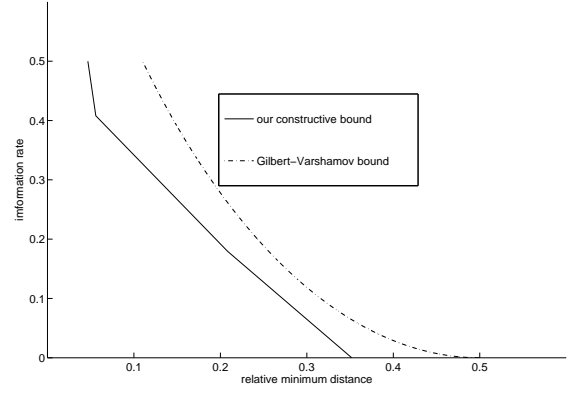}\\
  \caption{Asymptotic bound on self-orthogonal codes}\label{fig1}
\end{figure}

\section*{Acknowledgment}
The author is grateful to Profs. Keqin Feng, Jianlong Chen and
Chaoping Xing for their guidance.

\end{document}